\documentclass[12pt]{article}
\usepackage{latexsym,amsmath,amssymb,epsfig,graphicx}

\newcommand{\pkt}{\; .}
\newcommand{\kma}{\; ,}
\newcommand{\nn}{\nonumber}
\newcommand{\bea}{\begin{eqnarray}}
\newcommand{\eea}{\end{eqnarray}}
\newcommand{\be}{\begin{equation}}
\newcommand{\ee}{\end{equation}}
\newcommand{\beast}{\begin{eqnarray*}}
\newcommand{\eeast}{\end{eqnarray*}}
\newcommand{\eqn}[1]{(\ref{#1})}
\newcommand{\cald}{{\cal D}}
\newcommand{\call}{{\cal L}}
\newcommand{\calg}{{\cal G}}
\newcommand{\calm}{{\cal M}}
\newcommand{\caln}{{\cal N}}
\newcommand{\bfm}{{\bf M}}

\newcommand{\tr}{{\rm tr}}
\begin{document}
\begin{titlepage}
\begin{flushright}
DO-TH-04/11 \\
hep-th/0411162 \\
November 2004
\end{flushright}

\vspace{20mm}
\begin{center}
{\Large \bf
Self-consistent bounces in two dimensions}
\vspace{10mm}

{\large
J\"urgen Baacke \footnote{e-mail: baacke@physik.uni-dortmund.de}
and
Nina Kevlishvili  
\footnote{e-mail: nina.kevlishvili@het.physik.uni-dortmund.de}}

\vspace{15mm}

{\large  Institut f\"ur Physik, Universit\"at Dortmund \\
D - 44221 Dortmund, Germany
}

\vspace{15mm}

\bf{Abstract}
\end{center}
We compute bounce solutions describing false vacuum decay
in a $\Phi^4$ model in two dimensions in the Hartree approximation,
thus going beyond the usual one-loop corrections to the decay rate.
We use zero energy mode functions of the fluctuation operator
for the numerical computation of the
functional determinant and the Green's function. We thus avoid the
necessity of discretizing the spectrum, as it is necessary when one uses
numerical techniques based on eigenfunctions. Regularization is performed
in analogy of standard perturbation theory; the renormalization
of the Hartree approximation is based on the 
two-particle point-irreducible (2PPI) scheme.
The iteration towards the
self-consistent solution is found to converge for some range of the
parameters. Within this range we find the corrections to
the leading one-loop approximation to be relatively small,
not exceeding one order of magnitude in the total transition
rate.
\end{titlepage}

\setcounter{page}{2}

\section{Introduction}
\label{intro}
One of the possible mechanisms at work in the cosmology
of the early universe is the transitions between 
two different ground states, a metastable one in which the universe
may be trapped, and a stable one, the true vacuum
\cite{Kobzarev:1974cp,Coleman:1977py,Callan:1977pt,Coleman:1980aw,Linde:1981zj}
As in an infinite volume system
 a vacuum tunnelling or thermal over-the-barrier
transition cannot occur globally, such processes occur
by the spontaneous formation of regions of ``true vacuum'' inside
the metastable state. The transitions may occur either by quantum
tunnelling or by thermal over-the-barrier transitions.
If they occur by vacuum tunnelling in a space of infinite
spatial extension, the classical
solution which describes the local tunnelling process is called
``the bounce''. For a $D$-dimensional space-time the solution
is a solution in $D$-dimensional Euclidean space, which is
$SO(D)$ symmetric. 
Here we will consider $(1+1)$-dimensional space time, and, therefore,
a bounce in $2$ Euclidean dimensions.  

In leading order the transition amplitude is given by
\cite{Langer:1969bc,Langer:1967ax,Affleck:1980ac,Coleman85}
\be \label{Gammaoneloop}
\Gamma^{1-\rm{loop}}=\frac{S_{\rm cl}}{2\pi}\cald^{-1/2}
\exp(-S_{\rm cl})
=\frac{S_{\rm cl}}{2\pi}\exp(-S_{\rm cl}-S_{\rm 1-l})
\pkt\ee
where $S_{\rm cl}$ and $S_{\rm 1-l}$ are the classical and
one-loop effective actions, respectively.
The prefactor arises from the functional integration over the translation mode
$\eta_0=\caln \vec\nabla \phi$, where $\phi$ is the classical solution.
By a virial theorem the normalization of this mode is related to
the classical action via
\be \label{zeronorm}
\caln^{-2}=S_{\rm cl}
\pkt\ee

The computation of the classical action and of the one-loop
quantum corrections has been performed by several groups 
for various models \cite{Baacke:1993ne,Baacke:1994bk,
Baacke:1995bw,Strumia:1998qq,Munster:1999hr,Munster:2000kk}
The quantum action is of course
divergent. Its renormalization should and can be done in exact analogy
to the standard renormalized perturbation theory. This has been
been emphasized in particular in Refs. \cite{Baacke:1991nh,
Baacke:1993ne,Graham:2002fi,Graham:2002xq,Graham:2002fw}
where in this, or related, contexts the Born series expansion
has been used to separated the divergent parts in a Lorentz-covariant
way.  

One may ask whether this leading-order formula, which is based
on the semiclassical expansion, will receive strong corrections
in higher orders. This will of course depend on the precise
system and on the parameters. For electroweak bubble nucleation,
a thermal over-the-barrier transition, these corrections were
found to be strong \cite{Baacke:1995bw}, for the parameter 
sets which at the time
were believed to be realistic, and next-to-leading
order corrections have been considered by S\"urig \cite{Surig:1997ne}.  
What he computed
were bubble solutions which minimized not the classical but the
entire one-loop effective action. This is different from
other calculations \cite{Bodeker:1993kj}
in which the bubble was calculated
using the effective {\em potential} and quantum-corrected kinetic 
terms. S\"urig's calculations
were performed for a coupled channel problem involving
Higgs and gauge fields, which somewhat conceals the simplicity
and technical elegance of the approach, demonstrating on the other hand
that it can be used for such involved and realistic problems.
More recently the problem of computing self-consistently 
corrections beyond one-loop
has been taken up by Bergner and Bettencourt \cite{Bergner:2003id}
who use the Cornwall-Jackiw-Tomboulis (CJT)\cite{Cornwall:1974vz}
 or two-particle-irreducible
(2PI) approach. They include the first nonleading loop correction,
which results in the Hartree approximation. The same approximation
also occurs as the lowest nontrivial loop correction in the
so-called two-particle-point-irreducible (2PPI) approach of
Coppens and Verschelde \cite{Verschelde:1992bs,Coppens:1993zc}. 
In comparison the S\"urig's approach, the Hartree
approximation takes into account the back reaction of the quanta
not only to the classical field, but also to the quantum fluctuations
themselves. Bergner and Bettencourt have considered such 
self-consistent corrections for two systems: the kink one
space dimension
and the bounce solutions in two Euclidean dimensions. 
Here we will reconsider their
approach, but use a different method for computing the
quantum corrections, analogous to S\"urig's, but generalized to the
Hartree approximation. Both the Green's functions and the 
fluctuation determinant will be computed using mode functions
of the fluctuation operator at zero frequeny, thus avoiding
a cumbersome summation over eigenmodes. Unlike
Bergner and Bettencourt we find, within our parameter range,
that the fluctuation operator has an unstable mode and
a mode at almost zero frequency which can be identified
as translation mode, both of which then can be dealt with in analogy 
to the one-loop case \cite{Baacke:1993ne,Baacke:2003uw}. For the
parameter values of Ref. \cite{Bergner:2003id} our iterative
computation of the self-consistent configuration does
not converge, so that a direct comparison is not
possible.

While the computation of the corrections to vacuum tunnelling
in the Hartree approximation is new in Ref. \cite{Bergner:2003id} 
the  Hartree approximation has been used to a great extent
in finding selfconsistent fermionic lumps describing the nucleon
in the chiral quark model \cite{Christov:1995vm,Alkofer:1994ph}.
There techniques similar to the ones discussed  here can and have been applied
\cite{Baacke:1998nm,Baacke:2000bv}   

The plan of the paper is as follows: in section 
\ref{thebounce} we recall the basic
formulae of the classical bounce solution. In section 
\ref{Hartreebounce} we formulate
the Hartree approximation, on the basis of the 2PPI formalism. In section
\ref{translation} we discuss the translation mode problem. In section 
\ref{greensfunction}
we explain our way of computing the Green's function, in 
section \ref{flucdet} we describe the computation of the
fluctuation determinant. The numerical aspects of
the unstable and translation modes are presented 
in section \ref{unstableandtranslation}. Renormalization
is discussed in section \ref{renormalization}.
Our numerical results are the subject of section 
\ref{numericalresults}. Conclusions
and an outlook are presented in section \ref{conclusion}.


\section{The bounce}
\setcounter{equation}{0}
\label{thebounce}
Let us consider a scalar field theory in 2D, with the Lagrange
density
\be
\call=\frac{1}{2} \partial_\mu \Phi \partial^\mu \Phi -U(\Phi)\kma
\ee
where the  potential $U(\Phi)$ is given by
\begin{equation}
U(\Phi)=\frac{1}{2} m^2\Phi^2-\eta\Phi^3
+\frac{1}{8}\lambda\Phi^4 \kma
\end{equation}
and displays two minima, one at $\Phi=0$, and the other one
at $\Phi=\Phi_->0$. The value of the potential at this second minimum
is lower and represents the ``true vacuum'', while $\Phi=0$ 
represents the ``false vacuum''.

The bounce is an SO(2) symmetric classical solution of the
Euclidean field equations ($t\to -i\tau$). We denote the
Euclidean variables by $x_1=x$ and $x_2=\tau$. The radius is
$r=\sqrt{x_1^2+x_2^2}$, and the classical field is
denoted as $\phi(r)$.
 The classical Euclidean action is given by
\be
S_{\rm cl}[\phi]=\int d^2 x \left[\frac{1}{2}\left(\nabla \phi\right)^2
+U(\phi)\right]\kma
\ee
and the bounce which minimizes this action satisfies
\be
-\Delta_2\phi +U'(\phi)=0\kma
\ee
or
\be
-\frac{d^2\phi}{dr^2}-\frac{1}{r}\frac{d\phi}{dr} + U'(\phi)=0\kma
\ee
and boundary conditions
\begin{equation}
\frac{d\phi}{dr}|_{r=0}=0,\qquad \phi_{r\to\infty}=\Phi_+\pkt
\end{equation}

The one-loop correction to the classical action is given by
\be
S_{\rm 1-l}=\frac{1}{2}\ln {\det}' \frac{-\Delta_2+U''(\phi(\vec x))}
{-\Delta_2+m^2}=\frac{1}{2}\ln \cald[\phi]\kma
\ee
where $m$ is the mass in the false vacuum,
\be
m^2=U''(0)
\pkt
\ee
and where the prime denotes that the translation zero mode is removed
and that one replaces the imaginary frequency of the unstable mode
by its absolute value.

The transition rate from the false to the true vacuum  is given,
in the one-loop approximation, by
\be
\Gamma^{1-\rm{loop}}=\left(\frac{S_{\rm cl}}{2\pi}\right)\cald^{-1/2}
\exp(-S_{\rm cl})
=\left(\frac{S_{\rm cl}}{2\pi}\right)\exp(-S_{\rm cl}-S_{\rm 1-l})
\pkt\ee

The prefactor arises originally as the normalization of the
zero mode. The presence of a zero mode in this approximation 
is demonstrated by taking the gradient of
the classical equation of motion: 
\be
\nabla_i \left[-\Delta_2\phi +U'(\phi)\right]=
\left[-\Delta_2  +U''(\phi)\right]\nabla_i \phi=0\pkt
\ee
Its normalization, defined by $\eta_{0i}=\caln_0\nabla_i\phi$,
and the condition that $\eta_{0i}$ is normalized to unity,
is given by
\be
\caln_0^{-2}=\int d^2 x \left(\nabla_i\phi\right)^2
\kma\ee 
where there is no summation over $i$.
The right hand side is, for a spherically symmetric
solution $\phi(r)$, equal to the kinetic term.
Furthermore, one finds that in two dimensions, using a
scaling argument, that $\int d^2x U(\phi)=0$ when 
$\phi$ is the classical solution. We therefore obtain
$\caln_0^{-2}=S_{\rm cl}$, i.e., Eq. \eqn{zeronorm}.


\section{The bounce in the Hartree approximation}
\setcounter{equation}{0}
\label{Hartreebounce}

The Hartree approximation can be derived from the 2PPI formalism
as follows:
The effective action in this formalism is given by
\be\label{effac}
S_{\rm eff}[\calm^2,\phi]=
S_{\rm cl}[\phi]+\Gamma^{2PPI}[\calm^2,\phi]
-\frac{3\lambda}{8}\int d^2x \Delta^2(\vec x)
\kma\ee
up to renormalization counter terms discussed in section 8.
Here $\Delta$ is  a local insertion into the
propagator which has the form
\be
\calg^{-1}(\vec x)=-\Delta_2 +\calm^2(\vec x)\kma
\ee
with the definition
\be \label{calmdef}
\calm^2=m^2+\frac{3}{2}\lambda\phi^2-6\eta \phi
+\frac{3}{2}\lambda \Delta
\pkt \ee 
$i
\Delta$ itself is defined by the gap equation
\be
\frac{1}{2}\Delta(\vec x)=\frac{\delta}{\delta \calm^2(\vec x)}
\Gamma^{2PPI} 
\pkt \ee
Finally $\Gamma^{2PPI}$ is the sum of all two particle point
irreducible graphs, in which all internal propagators
have the effective masses $\calm^2$. 
A graph is two particle point reducible (2PPR)
if it falls apart if {\em two lines meeting at a point} are cut.
To lowest order in a  loop expansion $\Gamma^{2PPI}$
is given by a simple loop, i.e.
\be
\Gamma^{2PPI}=\frac{1}{2}\ln\det\frac{-\Delta_2+\calm^2}
{-\Delta_2+m^2}
\pkt\ee
We will come to the zero mode question later.
In this approximation, $\Delta$ is given by
\be
\Delta(\vec x)=\frac{\delta \Gamma^{2PPI}}{\delta \calm^2(\vec x)}=
<\vec x|\frac{1}{-\Delta_2+\calm^2}|\vec x>=\calg(\vec x,\vec x)
\pkt \ee
Here the Green's function $\calg$ is defined by
\be
(-\Delta_2+\calm^2)\calg(\vec x,\vec y)=\delta^2(\vec x-\vec y)
\pkt\ee
In taking variational derivatives of the effective action
we have to consider $\Delta$ as a function of $\calm^2$
and $\phi$, i.e., in the last term
of Eq. \eqn{effac} we have to replace
\be
\Delta=-\phi^2+\frac{2}{3\lambda}\left(\calm^2-m^2+6\eta\phi\right)\kma
\ee
by Eq. \eqn{calmdef}.

Taking the variational derivative of the effective action 
with respect to the field $\phi$ then leads to
\be
-\Delta_2\phi
+U'(\phi(\vec x))+ \frac{3}{2}\calg(\vec x,\vec x)
\left[\lambda \phi(\vec x)-2\eta\right]=0
\pkt
\ee
Using rotational symmetry we obtain explicitly
\bea\nn
&&-\frac{d^2\phi(r)}{dr^2}-\frac{1}{r}\frac{d\phi(r)}{dr}
+m^2\phi(r)-3 \eta\phi^2(r)+
\frac{\lambda}{2}\phi^3(r) 
\\\nn
&&\hspace{10mm}+ \frac{3}{2}\calg(\vec x,\vec x)\biggl|_{|\vec x|=r}
\left[\lambda \phi(r)-2\eta\right]=0
\pkt
\eea
The back reaction of the quantum modes onto themselves is contained
in $\calm^2(\vec x)$.


\section{The translation mode}
\label{translation}\setcounter{equation}{0}
In the formula for the transition rate the functional determinant
was modified by removing the zero mode, whose functional
integration is not Gaussian and which leads to the prefactor
$S_{\rm cl}/2\pi$. Once we modify the fluctuation operator
by introducing the quantum back reaction, there will be
no zero mode anymore. 
However, if $\calg(\vec x,\vec y)$ were the exact Green's
function
\be
\calg_{\rm exact}(\vec x,\vec y)=\frac{\delta^2 \Gamma_{\rm eff}}
{\delta \phi(\vec x)\delta \phi(\vec y)}\kma
\ee
the presence of the zero mode would still follow 
from the classical equation of motion
$\delta \Gamma_{\rm eff}/\delta \phi(\vec x)=0$
and translation invariance. As an artefact of the approximation
the zero mode is shifted to some finite value $\omega_t^2\neq 0$.
This is a well known problem, which also arises in other
applications. So in the Hartree approximation
to the $O(4)$ sigma model, the pions do not have
mass zero, in spite of their role as Goldstone bosons
\cite{Okopinska:1995su,Chiku:1998kd,Nemoto:1999qf}  
As long as the corrections are
small, the lowest eigenvalue of the fluctuation operator
in the partial wave $l=1$ will still be close to zero, 
as found by Surig in his calculations
on bubble nucleation. Of course, if we do not remove this
mode, the Green's function will receive a huge contribution
from it. Furthermore, removing a factor of dimension $(energy)^4$ 
from the functional determinant is 
necessary in order to provide the transition
probability with the correct dimension $(energy)^2$.

We take here a pragmatic point of view in removing the
``almost zero'' mode from the determinant. 
The formalism then requires to remove the translation
mode contribution from the Green's function as well, as 
$\Delta$ is now the functional derivative of the modified functional
determinant. 

Of course identifying and removing the ``would be''
translation mode in this way  constitutes an approximation
which comes in addition to the Hartree approximation itself. 
We will not be able to trust this approach if either of these
approximations leads to large modifications of the
bounce and of its determinant. In the case of the 
zero mode we will expect $\omega_t^2$ to satisfy
$|\omega_t^2|<< \Delta E$ where $\Delta E$ is the typical
level spacing. As most of the modes are continuum modes,
the ``typical'' level spacing would be the energy
difference between the zero mode and the unstable mode.
We will continue this discussion in section 9 in conjunction
to our numerical results. The technical problem
of removing the pole will be dealt with in the section 8.

A more fundamental approach has been taken
long ago, when the mass of the sine-Gordon soliton
received great attention. Various schemes 
\cite{Tomboulis:1975gf,Christ:1975wt,Gervais:1974dc,Gervais:1975pa,Baacke:1976gy,Baacke:1977pq} were formulated
to deal with the problem of translation invariance and
the projection to zero momentum states. These have been used
in order to compute the two-loop correction to the
soliton mass \cite{deVega:1976sm,Verwaest:1977tw}. 
They become very involved in higher loop
orders, and we are not aware of a formulation for the
case of a resummation as the one considered here.
We therefore have not followed such an approach.

As mentioned above, the prefactor $S_{\rm cl}/2\pi$
originates from integrating out the zero mode, and by using 
virial theorem relating the normalization of the zero mode
to the classical action. In the Hartree approximation this virial theorem
no longer holds. Nevertheless from what we have discussed above,
the exact zero mode should be $\nabla \phi_{\rm cl}$, where
$\phi_{\rm cl}$ now is the self-consistent profile. We therefore will
use its normalization as the prefactor, so that the transition  
rate in the Hartree approximation becomes
\be \label{GammaHartree}
\Gamma^{\rm Hartree}=\frac{\int d^2x \left(\nabla\phi_{\rm cl}\right)^2}{2\pi}
e^{\displaystyle -S_{\rm eff}}\kma
\ee 
where the effective action is defined in Eq. \eqn{effac}. 


\section{Computation of the Green's Function}
\label{greensfunction}
\setcounter{equation}{0}

In order to include the back-reaction of the quantum fluctuations
to the bounce in the Hartree approximation we need
the Green's function $\calg(\vec x,\vec x')$ of the
(new) fluctuation operator. In fact the Green's function
is usually discussed in a more general form, as a function
of energy. Here such a concept corresponds to introducing
an additional third dimension. We will choose it spacelike,
thus introducing an Euclidean time.
Using translation
invariance in the time direction we introduce the Fourier
transform $\calg(\vec x,\vec x',\nu^2)$, where $\nu$ is
the Euclidean frequency. The generalization, equivalent
of introducing an additional dimension, is necessary for discussing 
the translation mode and reappears in the formulation of the
determinant theorem in the next section.
The Green's function satisfies
\begin{equation}
[-\Delta_2+m^2+V(r)+\nu^2]\calg(\vec x,\vec x')=\delta^2(\vec x -\vec x')
 \kma
\end{equation}
with
\be
V(r)=-6\eta\phi(r)+\frac{3}{2}
\lambda \left(\phi^2(r)+\calg(\vec x,\vec x)\right)
\pkt\ee
The Green's function can be expressed by the eigenfunctions
of the fluctuation operator. We denote them by $\psi_\alpha(\vec x)$,
they satisfy
\be
[-\Delta_2+m^2+V(r)]\psi_\alpha(\vec x)=
\omega_\alpha^2 \psi_\alpha(\vec x)
\pkt\ee
In terms of these functions the Green's can be written as
\be\label{greenformal1}
\calg(\vec x,\vec x',\nu^2)=
\sum_\alpha\frac{\psi_\alpha(\vec x)\psi_\alpha(\vec x')}
{\omega_\alpha^2+\nu^2}
\pkt\ee
We may, furthermore, decompose the Hilbert space into angular momentum
subspaces, introducing eigenfunctions $\exp(il\varphi)R_{nl}(r)$,
where $\varphi$ is the polar angle and where the radial wave 
functions $R_{nl}(r)$ are eigenfunctions of
the partial wave fluctuation operator:
\be
\left[-\frac{d^2}{dr^2}-\frac{1}{r}\frac{d^2}{dr^2}+\frac{l^2}{r^2}+
m^2+V(r)\right]R_{nl}(r)=\omega_{nl}^2 R_{nl}(r)
\pkt\ee
Then the Green's  function takes the form
\be\label{greenformal2}
\calg(\vec x,\vec x')=\sum_l\sum_n e^{il(\varphi-\varphi')}
\frac{R_{nl}(r)R_{nl}(r')}{\omega_{nl}^2+\nu^2}
\pkt \ee
These expressions are formal, we have discrete and continuum states,
so the sum includes summation over discrete states and integration
over continuum states. While these expressions are very suitable
for discussions on the formal level, they are not very suitable for numerical
computation. In particular, if one uses these expressions for
the numerical computation, it becomes necessary to discretize
the continuum states by introducing a finite spatial boundary.

There is a well known alternative way of expressing Green's functions.
Consider first the free Green's function obtained for $V(r)=0$.
It can be written as
\be
G_0(\vec x,\vec x',\nu^2)=\int{\frac{d^2k}{(2\pi)^2}\frac{e^{i\vec k\cdot
(\vec x-\vec x')}}{k^2+m^2+\nu^2}}\kma
\ee
and this may be expanded as
\be
G_0(\vec x,\vec x',\nu^2)=
\frac{1}{2\pi}\sum^\infty_{l=-\infty}
e^{il(\varphi-\varphi')}I_l(\kappa r_<)K_l(\kappa r_>)\kma
\end{equation}
where $r_< = \min {|\vec x|,|\vec x'|}$,
$r_> = \max {|\vec x|,|\vec x'|}$ and $\kappa^2=m^2+\nu^2$.
Note that ultimately $\omega^2$ will be zero and
$\kappa=m$, so we have to deal with
the modified Bessel functions $I_l$ and $K_l$.
They satisfy
\be
\left[-\frac{d^2}{dr^2}-\frac{1}{r}\frac{d^2}{dr^2}+\frac{l^2}{r^2}+
\kappa^2\right]B_l(\kappa r)=0\kma
\ee
where $B_l$ stands for $I_l$ or $K_l$. $I_l(\kappa r)$ is
regular at $r=0$ and $K_l(\kappa r)$ is exponentially decreasing
for $r\to \infty$. Their Wronskian is given by
\be \nonumber
K_l(\kappa r)dI_l(\kappa r)/dr-I_l(\kappa r)dK_l(\kappa r)/dr=1/r
\pkt\ee
We now expand the exact Green's function in an analogous way
by  the ansatz
\be \label{Green}
\calg(\vec x,\vec x',\nu^2)=
\frac{1}{2\pi}\sum_{l=-\infty}^\infty 
e^{il(\varphi-\varphi')}f_l^-(r_<,\nu^2)f_l^+(r_>,\nu^2)
\pkt\ee
The functions $f_l^{\pm}(r,\nu^2)$ 
satisfy the mode equations
\be
\left[-\frac{d^2}{dr^2}-\frac{1}{r}\frac{d^2}{dr^2}+\frac{l^2}{r^2}+
\kappa^2+V(r)\right]f_l^{\pm}(r,\nu^2)=0\kma
\ee
and, furthermore, the following boundary conditions:
\be
\begin{array}{ll}
f_l^-(r,\nu^2)\propto r^l & r\to 0
\\
f_l^+(r,\nu^2)\propto \exp(-\kappa r)/\sqrt{
\kappa r} & r \to \infty
\end{array}\pkt
\ee
So $f_l^-$ is regular at $r=0$ and $f_l^+$ is
regular, i.e., bounded, as $r\to \infty$.
For $V(r)=0$ these boundary conditions are those satisfied by
$I_l(\kappa r)$ and $K_l(\kappa r)$, respectively.  
Furthermore, as the behaviour
at $r=0$ is determined by the centrifugal barrier, and
the behaviour for $r\to \infty$ by the mass term, these boundary conditions
are independent of the potential. If we write
\bea
f_l^-(r)&=&I_l(\kappa r)[1+h_l^-(r,\nu^2)]\kma
\\
f_l^+(r)&=&K_l(\kappa r)[1+h_l^+(r,\nu^2)]\kma
\eea
then the functions $h^\pm_l(r,\nu^2)$ become constant as $r\to 0$ and 
as $r\to \infty$, and for finite $r$ they interpolate smoothly
between these asymptotic constants.
If we impose the boundary conditions $h^\pm(r,\nu^2)\to 0$ 
for $r\to \infty$ then the Wronskian of $f_l^+$ and 
$f_l^-$ becomes identical to the one between $K_l(\kappa r)$ and
$I_l(\kappa r)$, i.e.,  equal to $1/r$.
Applying the fluctuation operator to our ansatz, Eq. \eqn{Green},
we then find
\be
\left[-\Delta_2+\kappa^2+V(r)+\right]G(\vec x,\vec x',\nu^2)=
\frac{1}{2\pi}
\frac{1}{r}\delta(r-r')\sum_{l=-\infty}^\infty e^{il(\varphi-\varphi')}
=\frac{1}{r}\delta(r-r')\delta(\varphi-\varphi')\pkt\ee
This completes the construction of the
Green's function.

Numerically we proceed as follows:
the functions $h_l^\pm$ satisfy
\bea
\{\frac{d^2}{dr^2}+[2\kappa\frac{I_{l}'(\kappa r)}{I_{l}(\kappa r)}
+\frac{1}{r}]\frac{d}{dr}\}h_l^-(r,\nu^2)
=V(r)[1+h_l^-(r,\nu^2)]\kma
\\
\{\frac{d^2}{dr^2}+[2\kappa\frac{K_{l}'(\kappa r)}{K_{l}(\kappa r)}
+\frac{1}{r}]\frac{d}{dr}\}h_l^+(r,\nu^2)
=V(r)[1+h_l^+(r,\nu^2)]\kma
\eea
which can be solved numerically. The second differential equation is solved
starting at large $r=\bar r$ with 
$h_l^+(\bar r,\nu^2)=h_l^{+'}(\bar r,\nu^2)=0$, and running
backward. $\bar r$ has to be chosen far outside the range of the
potential. In this region $h_l^\pm(r,\nu^2)$ are essentially constant.
For the first differential equation we first obtain a
solution $\tilde h_l(r,\nu^2)$ 
starting at $r=0$, with $\tilde h_l(0,\nu^2)=\tilde h_l'(0,\nu^2)=0$.
This function does not satisfy the boundary condition 
required for the Green's function, it will be used for the 
computation of the functional determinant. 
The function $h_l^-(r,\nu^2)$ is obtained from $\tilde h_l(r,\nu^2)$
via
\be\label{moderen}
h_l^-(r,\nu^2)=\frac{\tilde h_l(r,\nu^2)
-\tilde h_l(\bar r,\nu^2)}
{1+\tilde h_l(\bar r,\nu^2)}\kma
\ee
which obviously solves the differential equation with the appropriate
 boundary conditions.

Finally the Green's function is given by
\be \label{Green_h}
\calg(\vec x,\vec x',\nu^2)=
\frac{1}{2\pi}\sum_{l=-\infty}^{\infty} e^{il(\varphi-\varphi')}
I_l^-(\kappa r_<)
K_l^+(\kappa r_>)(1+h_l^-(r_<,\nu^2))(1+h_l^+(r_>,\nu^2)
\pkt\ee


\section{Computation of the Fluctuation Determinant}
\label{flucdet}
The fluctuation determinant which appears in the rate formula
\be
\cald={\det}'\frac{-\Delta_2+\calm^2}{-\Delta_2+m^2}\kma
\ee
can be written formally as an infinite product of eigenvalues
of the fluctuation operator. The prime denotes taking the absolute
value and removing the translation mode.
As in the previous section we introduce the
generalization
\be \label{def_Dtilde}
\tilde\cald(\nu^2)=
\det\frac{-\Delta_2+\calm^2+\nu^2}{-\Delta_2+m^2+\nu^2}
\pkt\ee
Note that we omit the prime, here.
Using the decomposition of the Hilbert space into angular momentum
subspaces we can write
\begin{equation}
\tilde\cald(\nu^2)
=\prod_{l,n}
\left[\frac{\omega^2_{ln}+\nu^2}
{{\omega^{{}2}_{l n{(0)}}+\nu^2}} \right]
=\prod_{l=0}^\infty \left[\frac{\det \bfm_l(\nu^2) }
{\det \bfm_l^{(0)}(\nu^2)}\right]^{d_l} \kma
\end{equation}
with the radial fluctuation operators
\be
\bfm_l(\nu^2)=-\frac{d^2}{dr^2}-\frac{1}{r}\frac{d}{dr}+\frac{l^2}{r^2}
+m^2+V(r)+\nu^2
\kma
\ee
as before. 
$d_l$ denotes the degeneracy. If we restrict $l$ to positive values then
$d_l=2$ for $l>0$ and $d_l=0$ for $l=0$.

According to a theorem on functional determinants of ordinary
differential operators 
\cite{Coleman85,Dashen:1974ci} we can express the ratios of the
partial wave determinants via

\begin{equation}
\frac{\det \bfm_l(\nu^2)}{\det \bfm_l^{(0)}(\nu^2) } =
\lim_{r\to\infty} \frac{\psi_l(\nu^2,r)}{\psi_l^{(0)}(\nu^2,r)} \kma
\end{equation}
where $\psi_l(\nu^2,r)$ and $\psi_l^{(0)}(\nu^2,r)$ 
are solutions to equations
\begin{equation}
\bfm_l(\nu^2)\psi_l(\nu^2,r)=0 \kma~~~~~~
\bfm_l^{(0)}(\nu^2)\psi_l^{(0)}(\nu^2,r)=0 \pkt
\end{equation}
with identical regular boundary conditions at $r=0$. Of course
\begin{equation}
\psi_l^{(0)}(\nu^2,r)=I_l(\kappa r) \pkt
\end{equation}
Furthermore we have
\begin{equation}
\psi_l(\nu^2,r)=\left[1+\tilde h_l(\nu^2,r)\right]I_l(\kappa r)\pkt
\end{equation}
where $\tilde h_l(\nu^2,r)$ is precisely the function we have
introduced in the previous section, except for the fact that we have
added $\nu^2$
to the fluctuation operator. We finally have
\begin{equation}
\frac{\det \bfm_l(\nu^2)}{\det \bfm_l^{(0)}(\nu^2) } = 
1+\tilde h_l(\nu^2,\infty) \kma
\end{equation}
and
\be
\ln \tilde \cald(\nu^2)=
\sum_{l=0}^\infty d_l\ln \left[1+\tilde h_l(\nu^2,\infty)\right]\pkt
\ee
The fluctuation determinant in the transition rate formula
and in the $2PPI$ formalism refers to the fluctuation
operators at $\nu^2=0$, and so in the numerical computation
we just need the functions $\tilde h_l(0,\infty)$, as for the Green's
function. The only exception is the translation mode we will
discuss in the next section.


\section{Unstable and translation modes}
\label{unstableandtranslation}
\setcounter{equation}{0}

In the one-loop formula for the transition rate the determinant
of the fluctuation operator appears as ${\det}' (-\Delta_2+\calm^2)$,
and the prime denotes two modifications with respect to the
naive determinant: 

(i) the unstable mode has an imaginary frequency,
corresponding to a negative eigenvalue $\omega_u^2=-\nu_u^2$ of the
fluctuation operator. It is to replaced by its absolute value.
This mode appears in the $s$-wave $l=0$ and manifests itself by
a negative value of $1+\tilde h_0(0,\infty)$. So here we have to take
the absolute value.

(ii) the translation mode manifests itself, in the one-loop
approximation, by the asymptotic limit $1+\tilde h_1(-\omega_t^2,\infty)=0$.
We denote the frequency of the translation mode, which is the lowest radial
mode in the $m=1$ partial wave, by $\omega_{10}=\omega_t$.
The fluctuation determinant
\eqn{def_Dtilde} has a factor $\omega_t^2+\nu^2$
which has to be removed, according to the definition
of ${\det}'$. Otherwise the logarithm of this expression, 
appearing in the functional
determinant, does not exist.  Furthermore the Green's function
is not defined either at $\nu^2=-\omega_t^2$. 
In the one-loop approximation $\omega_t^2=0$, in the
Hartree approximation it is close to zero, otherwise
we cannot trust our approximation of identifying this
mode with a ``would-be'' zero mode.

In the one-loop approximation the translation mode
is removed numerically in the following way\cite{Baacke:1993ne}: we
compute $\tilde h_1(\infty,\pm \epsilon^2)$ for some 
sufficiently small $\epsilon$ and replace
\be
\left[1+\tilde h_1(0,\infty)\right] 
 \to \frac{\tilde h_1(\epsilon^2,\infty)
-\tilde h_1(-\epsilon^2,\infty)}{2\epsilon^2}
\kma\ee
i.e., we take the numerical derivative at $\omega^2=0$.

As discussed above, beyond the one-loop approximation 
there is no exact translation
mode, but a pole in the Green's function appears very close
to $\nu^2=0$, and its wave function is close to
$\nabla \phi$. This contribution would make the corrections 
extremely large. As long as the corrections beyond one-loop
are small, the ``almost zero'' mode still corresponds
to a collective motion of the system, with almost vanishing
 restoring force. As discussed in section 3 we will continue
to remove it also in the Hartree approximation. We determine
the position of the eigenvalue by requiring
 $1+\tilde h_1(-\omega_t^2,\infty)$ to vanish, and compute the
numerical derivative not at $\nu^2=0$ 
but at $\nu^2=-\omega_t^2$,
i.e., we remove a factor $\omega_t^2+\nu^2$.

The Green's function in the $l=1$ channel
has, at $r=r'$, the form
\be
\calg_l(r,r,\nu^2)=\frac{R_t(r)^2}{\nu^2+\omega_t^2}
+\sum_{n\neq0}
\frac{R_{1,n}^2(r)}{\nu^2+\omega_{1n}^2}
\pkt
\ee
We can use the fact that the pole term is antisymmetric 
with respect to $\nu^2+\omega_t^2$ by computing 
the Green's function
at $\nu^2=-\omega_t^2\pm \epsilon^2$ and by 
taking the average of these two values. Then the pole term has
disappeared and the averaged Green's function takes the form
\be
\frac{1}{2}\left[\calg_1(r,r,-\omega_t^2+\epsilon^2)
+\calg_1(r,r,-\omega_t^2-\epsilon^2)\right]
=\sum_{n\neq 0}R_{1n}^2(r)\frac{\omega_{1n}^2-\omega_t^2}{
(\omega_{1n}^2-\omega_t^2)^2-\epsilon^4 }\pkt
\ee
As long as $\omega_t^2$ and $\epsilon^2$ are much smaller
than the $\omega_{1n}^2$ this is a good approximation to
the desired reduced Green's function 
\be
\left[\calg_1(r,r,0)\right]_{\rm red}=\sum_{n\neq 0} \frac{R_{1n}^2(r)}{
\omega_{1n}^2}\pkt
\ee

In the explicit numerical computation of the Green's function
 we use of course the  expression \eqn{Green}.
As evident from Eqs. (\ref{moderen}) the pole 
arises from the re-normalization of the mode function $\tilde h_1(\nu^2,r)$,
i.e., from dividing by $1+\tilde h_-(\nu^2,\infty)$.
In averaging over the Green's functions at $\nu^2=-\omega_t^2
\pm \epsilon^2$ we add two very large terms which almost cancel.
This can be done in a somewhat smoother way: if $\epsilon^2$
is sufficiently small we can assume that $1+\tilde h_1(\nu^2,\infty)$
passes through zero linearly and we may replace
\be
1+\tilde h_1(-\omega_t^2\pm\epsilon^2,\infty)
\to \pm\frac{1}{2}\left[\tilde h_1(-\omega_t^2+\epsilon^2,\infty)-
\tilde h_1(-\omega_t^2-\epsilon^2,\infty)\right]\pkt
\ee
The average over the Green's functions can then be cast into the form
\be
\left[\calg_1(r,r,0)\right]_{\rm red}\simeq
\frac{f_1^+(-\omega_t^2+\epsilon^2,r)\tilde f_1(-\omega_t^2+\epsilon^2,r)
-f_1^+(-\omega_t^2-\epsilon^2,r)\tilde f_1(-\omega_t^2-\epsilon^2,r)
}{\tilde h_1(-\omega_t^2+\epsilon^2,\infty)-
\tilde h_1(-\omega_t^2-\epsilon^2,\infty)}\pkt
\ee
where $\tilde f_1(\nu^2,r)=I_1(\kappa r)[1+\tilde h_1(\nu^2,r)]$
is the mode function $f_1^-$ before the re-normalization.


\section{Renormalization}
\label{renormalization}
In the previous sections we have presented the basic formalism and
its numerical implementation. There is still one point to be discussed:
divergences and renormalization. It is well known, that renormalization
in $(1+1)$ dimensions just requires normal ordering. This means in practice
that we have to redefine $\Delta=\calg(\vec x,\vec x)$ by
\be
\Delta(\vec x)=\calg(\vec x,\vec x)-\calg_0(\vec x,\vec x)
\kma\ee
or, in terms of the functions $h_l^\pm(r)$
\be
\Delta(\vec x)=\sum_{l=0}^\infty
d_l I_l(mr)K_l(mr)\left[h_l^-(r)+h_l^+(r)+h_l^-(r)h_l^+(r)\right]\pkt
\ee
Normal ordering is not unique, it depends on the mass used in the
free Green's function. We have used here the mass in the false
vacuum; this may be changed, but it essentially means that
we redefine our couplings $\eta$ and $\lambda$.

There is one more place where we have to subtract a tadpole
diagram: in the $\tr\log$ term. If we expand it perturbatively 
we find
\be
\left[\frac{1}{2}\ln \cald\right]_{\rm div}=
\frac{1}{2}\int d^2 x  V(r)\calg_0(\vec x,\vec x)\pkt
\ee
While it is trivial to remove this part in a perturbative
calculation it is less obvious how to remove it from
a nonperturbative one. Indeed we have computed $\ln \cald$
using the partial waves. However the divergent contribution
can be traced {\em exactly} in this partial wave representation:
it is given by the contributions of first order in $V(r)$, and these
can be computed exactly. 

We have
\be
\{\frac{d^2}{dr^2}+[2m\frac{I_{l}'(m r)}{I_{l}(m r)}
+\frac{1}{r}]\frac{d}{dr}\}\tilde h_l(r)
=V(r)[1+\tilde h_l(r)]
\pkt\ee
This inhomogeneous differential equation allows for an iterative
expansion of $\tilde h_l^-(r)$ and the first order contribution 
$h_l^{(1)}(r)$ is the solution of
\be
\{\frac{d^2}{dr^2}+[2m\frac{I_{l}'(m r)}{I_{l}(m r)}
+\frac{1}{r}]\frac{d}{dr}\}\tilde h_l^{(1)}(r)
=V(r)
\pkt\ee
This equation can easily be solved numerically. Then the renormalized
contribution of a partial wave $l$ to the fluctuation determinant
is given by
\be
J_l=d_l \left[\ln (1+\tilde h_l(\infty))-\tilde h_l^{(1)}(\infty)\right]
\pkt
\ee
This renormalization can be generalized to higher dimensions,
using the renormalization procedure of Verschelde \cite{Verschelde:2000dz}
which applies to the Hartree approximation. Then higher subtractions
are required, they can be likewise determined exactly within the
numerical procedure, see, e.g., \cite{Baacke:1993ne,Baacke:1991sa}.
While Verschelde' s procedure of removing the divergences is based on
an elaborate analysis of Feynman graph's, it can be, from the
practical point of view \cite{Nemoto:1999qf}, 
obtained by adding a counterterm
\be
S_{c.t.}=\int d^2 x\left[ A \calm^2(x) +\Lambda\right]
\pkt\ee
$\Lambda$ is the ``cosmological constant'' counterterm, in
$(3+1)$ dimensions there is a further counterterm
proportional to $\calm^4(x)$. Finiteness of the gap equation and
of the action require
\bea
A&=&-\frac{1}{8\pi}\left[\frac{2}{\epsilon}-\gamma+
\ln \frac{4\pi\mu^2}{m^2}\right]\kma 
\\
\Lambda&=&-m^2 A 
\pkt
\eea
This corresponds precisely to the subtractions we have made, 
which are identical to minimal subtraction. In
particular
\be 
\Delta(x)=\calg(x,x)+A={\rm finite}\pkt
\ee
\begin{figure}[htbp]
  \centering
\vspace{7mm}
   \includegraphics[scale=0.38]{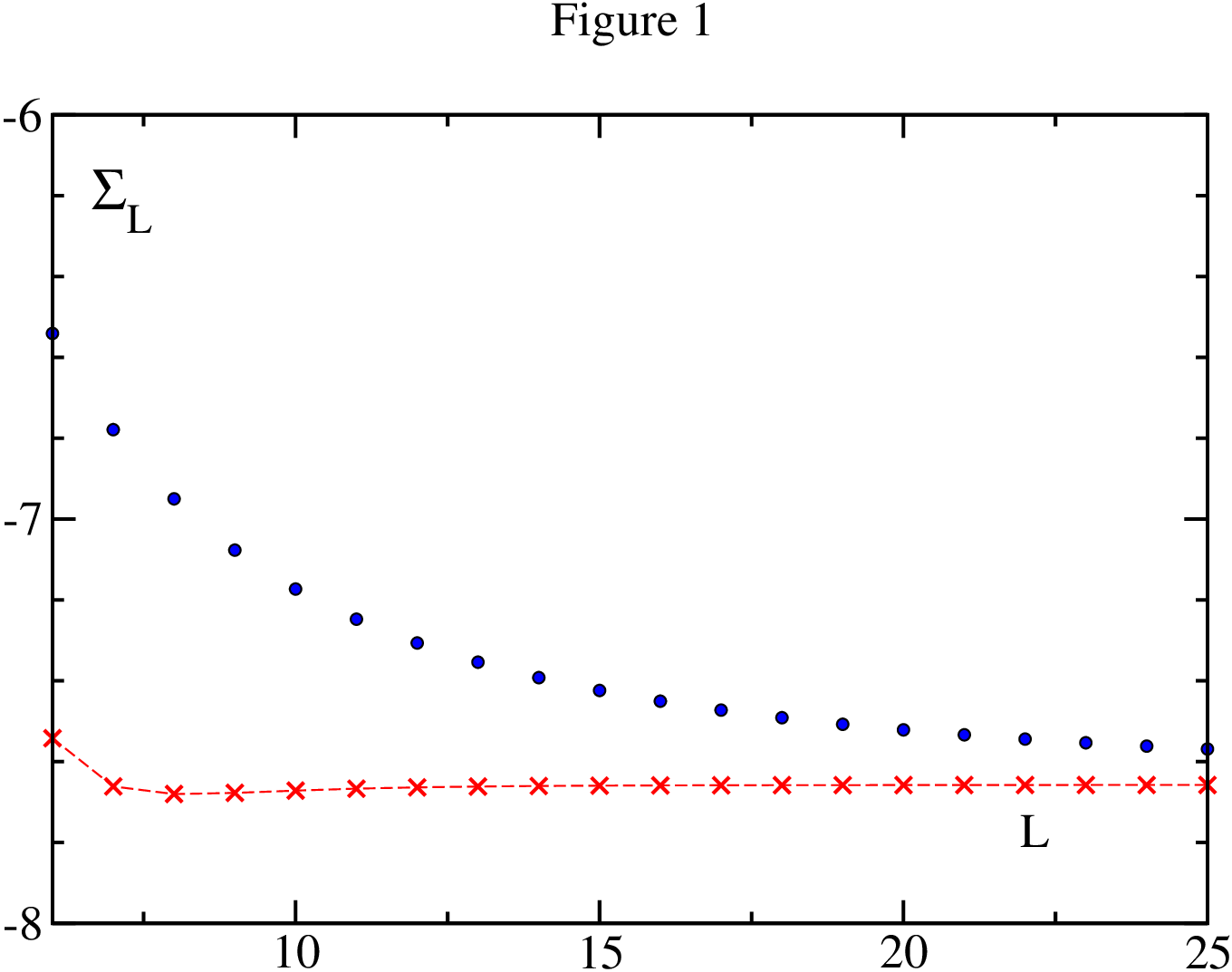}
\vspace{4mm}
  \caption{Convergence of the angular-momentum summation.
$\alpha=0.9$,$\beta=1$; dots: $\Sigma_L$; dashed line with
crosses: $\Sigma_L^{\rm as}$.}
  \label{fig:lconv}
\end{figure}
The expressions for the Green's function and
for the one-loop action imply summation over angular momentum 
$l$. After renormalization these converge in the expected form,
the single terms being proportional to $l^{-3}$, while the 
unsubtracted terms behave like $l^{-1}$. This has been verified
numerically and presents a cross check on the procedure and the
accuracy. We have appended the sum from $L=l_{\rm max}$ to
$\infty$ by adding this sum in its asymptotic form. We have used
a fit $A/l^3+B/l^4+C/l^5$ through the five values of $J_l$ at
$l_{\rm max}-5,\dots, l_{\rm max}$ in order to parameterize
the terms in this asymptotic sum.
In order to illustrate the convergence in $l$ we plot,
in Fig. \ref{fig:lconv}, the values $\Sigma_L$ and  of $\Sigma_L^{\rm as}$,
where the latter is the exact finite sum up to $L$ 
complemented by the asymptotic sum, for
$\lambda=0.9, \eta=0.5$. The convergence is
seen to be excellent. In our actual calculations we used
$l_{\rm max}=25$. The same procedure was applied to the
summation over angular momenta occuring in $G(\vec x,\vec x)$.


\section{Numerical results}
\label{numericalresults}
\setcounter{equation}{0}

The computational framework introduced in the previous sections
has been carried through for a representative part of the parameter space.
The parameter space is conveniently analyzed
\cite{Dine:1992wr,Baacke:1993ne} in terms
of two dimensionless variables $\alpha$ and $\beta$  which are introduced
as follows: We  define the
scaled variables: $X=mx$ and $\Phi=2 \eta/m^2 \phi$. Then the 
classical action takes the form
\bea\nonumber
S_{\rm cl}&=&\frac{m^4}{4 \eta^2}\int d^2 X\left[
\frac{1}{2}\left(\nabla_X\Phi\right)^2+\frac{1}{2}\Phi^2-
\frac{1}{2}\Phi^3+\frac{\alpha}{8}\Phi^4\right]
\\
&=& \beta\hat S_{\rm cl}(\alpha)\kma
\eea
with $\beta=m^4/4\eta^2$ and $\alpha=\lambda \beta/m^2$.
The variable $\alpha$ can take values 
$0<\alpha<1$ as a condition for the existence of the
classical bounce solution. $\alpha=1$ corresponds to
degenerate minima, the limit $\alpha\to 1$ is called the
thin-wall limit. This defines the parameter range for $\alpha$.
$\beta$ appears in front of the rescaled classical
action while there is no such factor in front 
of the quantum action which in the one-loop approximation only
depends on $\alpha$. Therefore
large $\beta$ imply a large classical action and therefore
relatively small quantum corrections, small values of $\beta$
make the classical action small and the quantum corrections 
relatively important. This consideration applies to the one-loop
approximation; for a self-consistent scheme the separation
of classical and quantum parts is not unique, but this consideration
may still serve as an estimate. Indeed we find that for $\beta \lesssim 0.8$
the Hartree iteration, even with an underrelaxation parameter, does not
converge. 

\begin{figure}[htbp]
  \centering
   \includegraphics[scale=0.28]{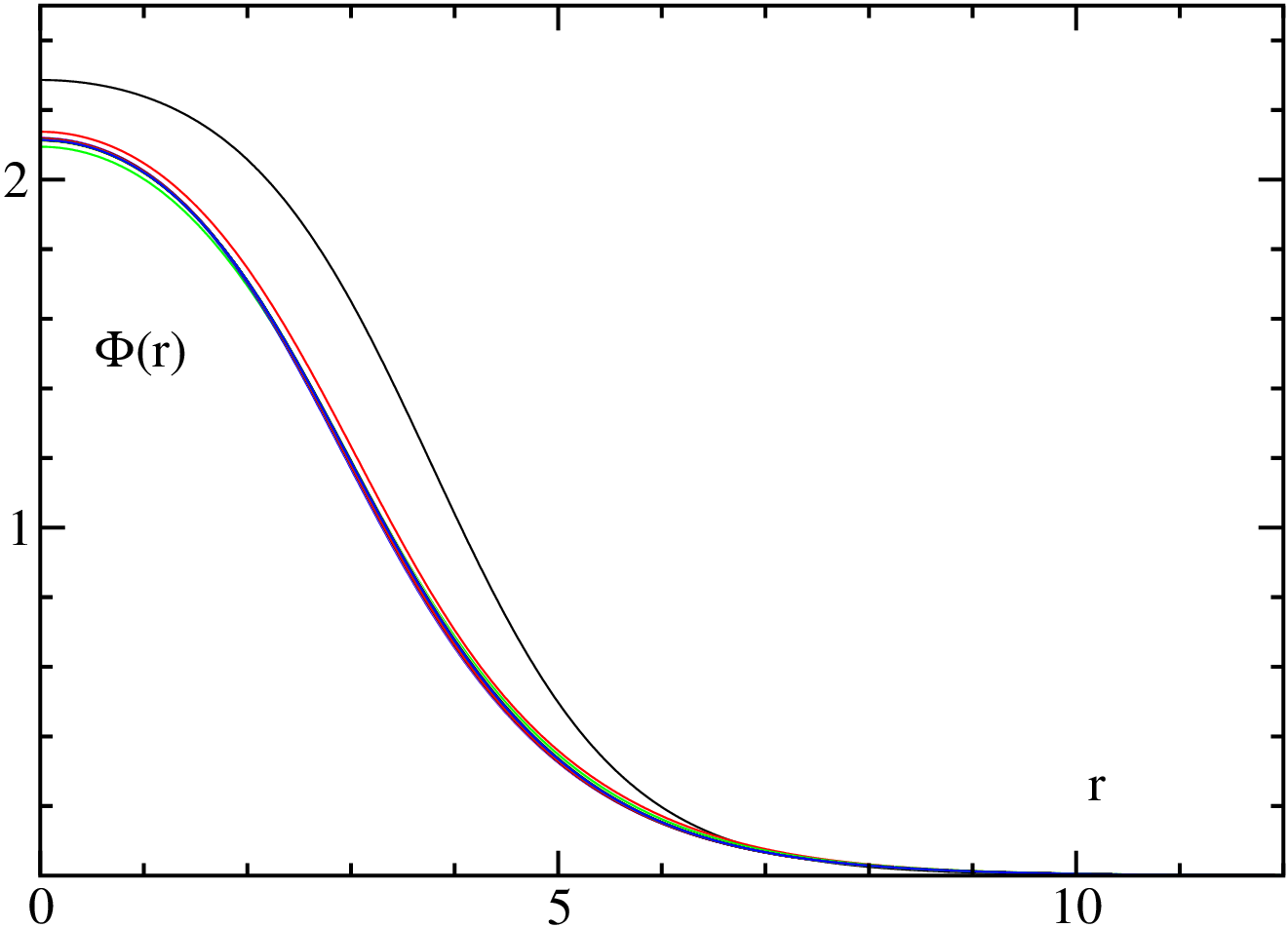}
\hspace{4mm}
   \includegraphics[scale=0.28]{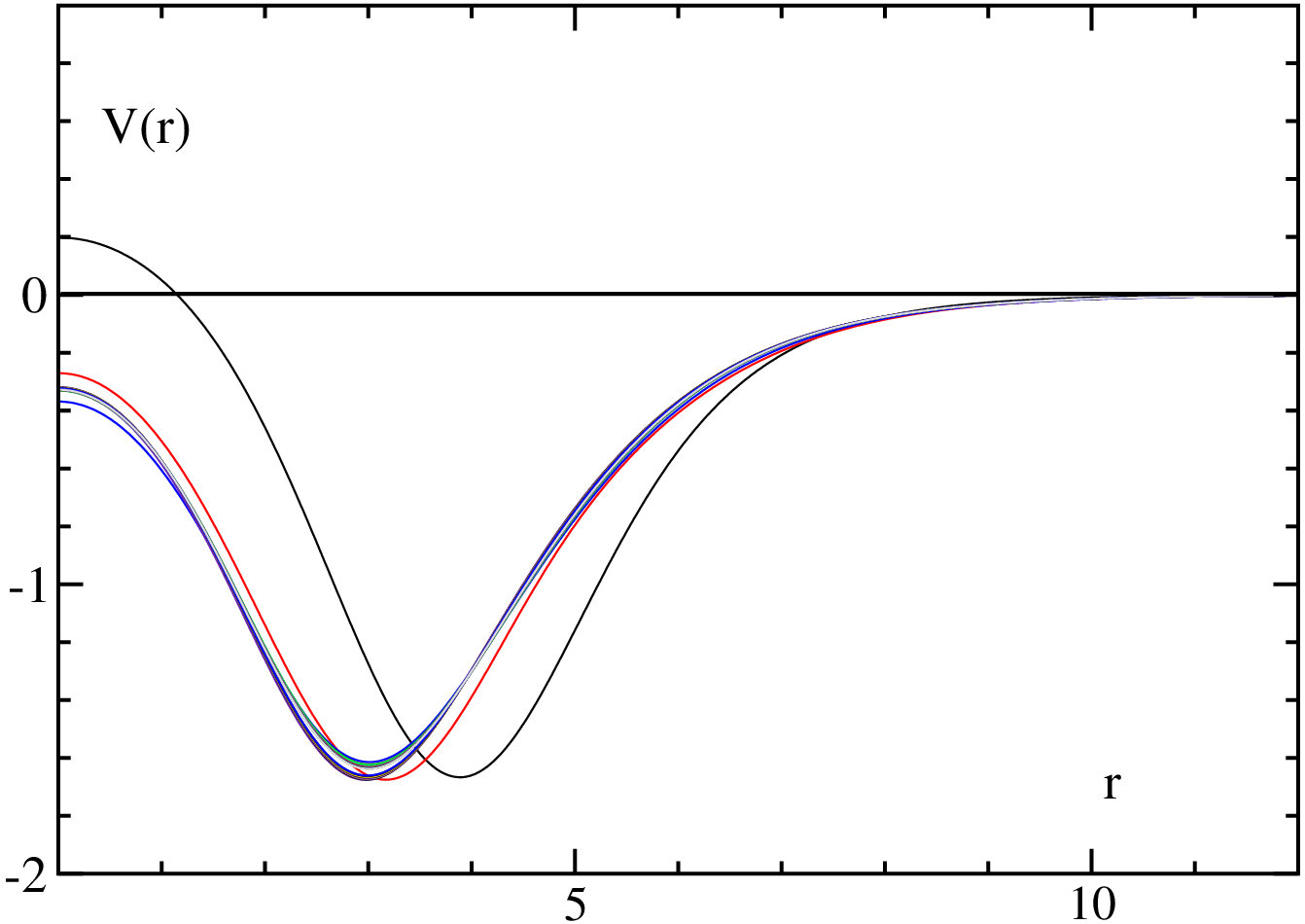}
\vspace{5mm}
  \caption{Convergence of the bounce profiles and of the
potential for $\alpha=0.9$ and $\beta=1$.}
  \label{fig:profconv}
\end{figure}

The computations were started with the one-loop approximation
and then iterated until the largest difference in the
profiles $\phi(r)$ between two subsequent
iterations $\max_r \Delta \phi(r)$ was smaller than $ 10^{-5}$.
The convergence of the profiles $\phi(r)$ and of the 
potential $V(r)$ is displayed in Figs. \ref{fig:profconv}(a) and (b), 
respectively,
for the case $\alpha=0.9,\beta=1$.
\vspace{6mm}

The iteration is found to converge for $\alpha \lesssim 0.9$
and $\beta \gtrsim 0.8$. The translation zero mode is found at
typically $|\nu^2|\simeq 10^{-4}$ in the one-loop approximation,
thus verifying to good accuracy the quality of the classical profile
and of the integration of the mode equation. In the Hartree
approximation the mode which we continue to identify with
the zero mode is located at values of $|\nu^2|\simeq 10^{-2}$,
the largest value of $0.1$ occurs for $\alpha=0.2,\beta=0.8$.
For $\alpha=0.9$ and $\beta=1$ the required accuracy
is obtained after $80$ iterations, this takes about $40$ seconds
with a $1.3$ GHz processor. For $\alpha < 0.9$ and $\beta > 1$ the 
convergence is much faster.

\begin{figure}[htbp]
  \centering
   \includegraphics[scale=0.28]{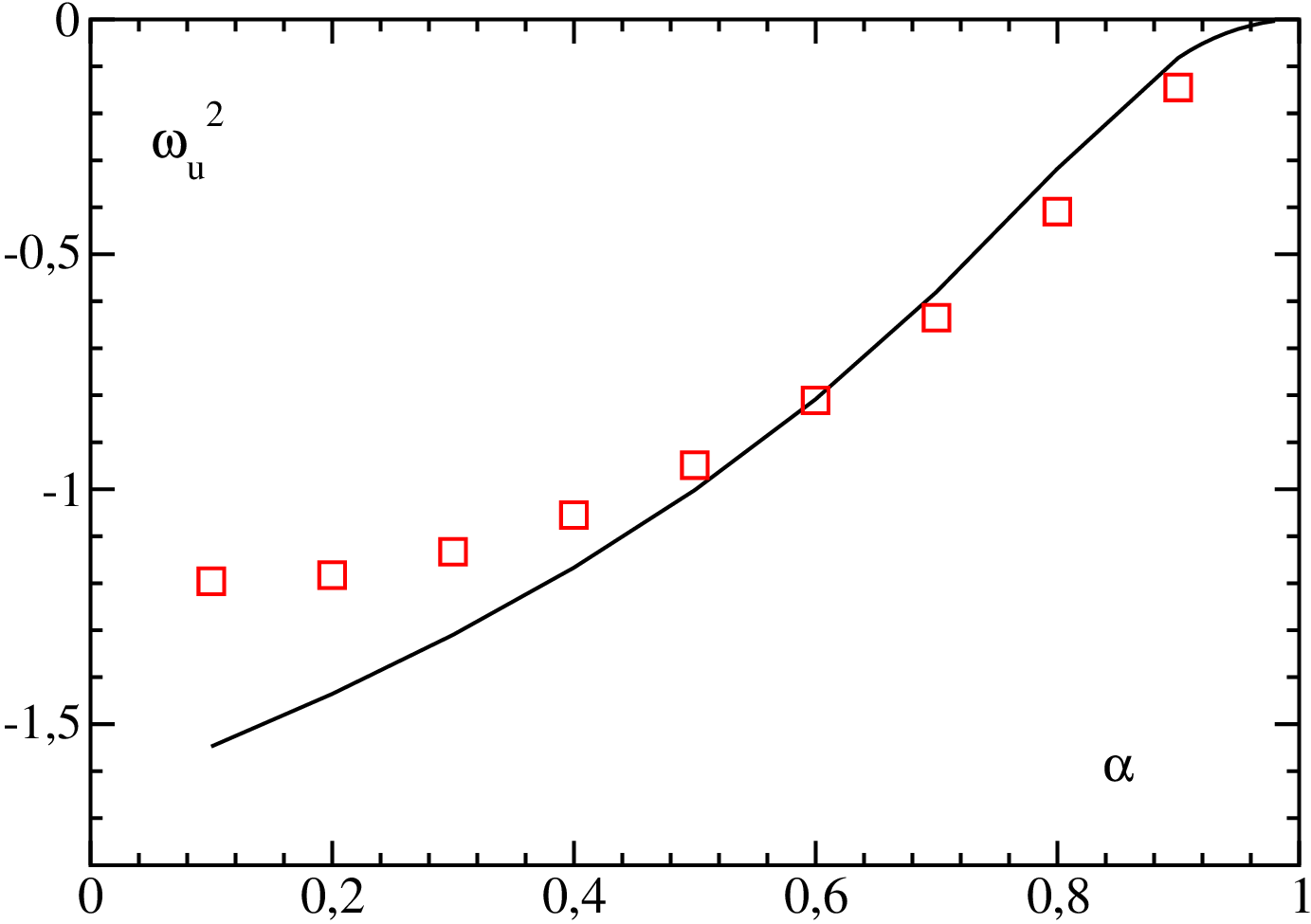}
\vspace{5mm}
  \caption{The unstable mode for $\beta=1$. Straight line: one-loop
approximation; squares: Hartree approximation.}
  \label{fig:unstable}
\end{figure}

\begin{figure}[htbp]
\centering
\includegraphics[scale=0.28]{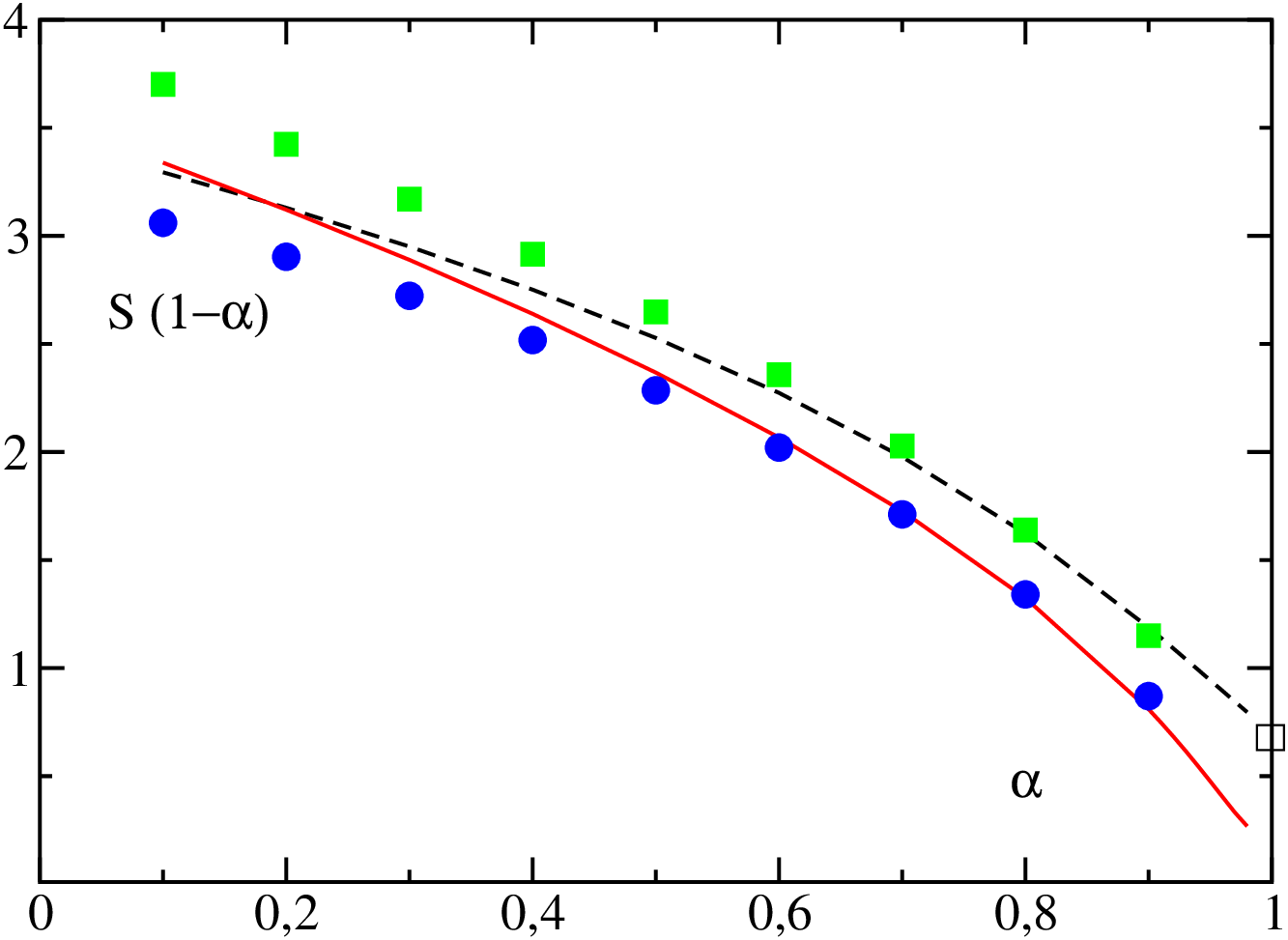}
\\\vspace{5mm}
  \caption{Classical and total actions as functions
of $\alpha$, multiplied by $(1-\alpha)$. 
Dashed line: $S_{\rm cl}^{1-{\rm loop}}$;
empty square: thin  wall limit of  $S_{\rm cl}^{1-{\rm loop}}$;
straight line:  $S_{\rm tot}^{1-{\rm loop}}$; full squares:
 $S_{\rm cl}^{\rm Hartree}$;
 full circles:  $S_{\rm tot}^{\rm Hartree}$.} 
\label{fig:alphadependence}
\end{figure}

The $\alpha$-dependence of the unstable mode in the
one-loop approximation is displayed in Fig. \ref{fig:unstable}.
One sees that it approaches zero for $\alpha\to 1$. This
is obviously the cause of the lack of convergence for $\alpha > 0.9$.
During the iteration towards the self-consistent Hartree solution,
the mode moves between positive and negative values
of $\nu^2$ while giving large contributions to $\tr \calg$.
If we look, see Fig. \ref{fig:alphadependence}, 
at the effective actions found in the Hartree
approximation then there is no evidence of any 
singular behaviour near $\alpha=0.9$. So we think that
the lack of convergence is just a technical problem which could
be overcome, but this certainly would need some new 
analytic idea, not just plain numeric efforts.
\vspace{6mm}

The dependence on the parameter $\alpha$ at fixed $\beta$ is displayed
in Fig. \ref{fig:alphadependence}. The classical actions and the
quantum actions in the one-loop and in the Hartree approximations
are plotted including a factor $(1-\alpha)$. Indeed all of them
are singular in the thin-wall limit $\alpha\to 1$. For the classical
action in the one-loop approximation the behavior can be
determined analytically using the standard technique (see e.g.
\cite{Coleman85}) as 
\be\label{thinwall}
\lim_{\alpha\to 1}S_{\rm cl}^{1-{\rm loop}}(1-\alpha)=
\beta\frac{2\pi}{9}
\pkt \ee 
The one loop results extend to $\alpha=0.97$,
for the Hartree results we were not able to obtain convergence
beyond $\alpha\gtrsim 0.9$ as already mentioned. 

\begin{figure}[htbp]
\centering
(a)\hspace{10mm}\includegraphics[scale=0.28]{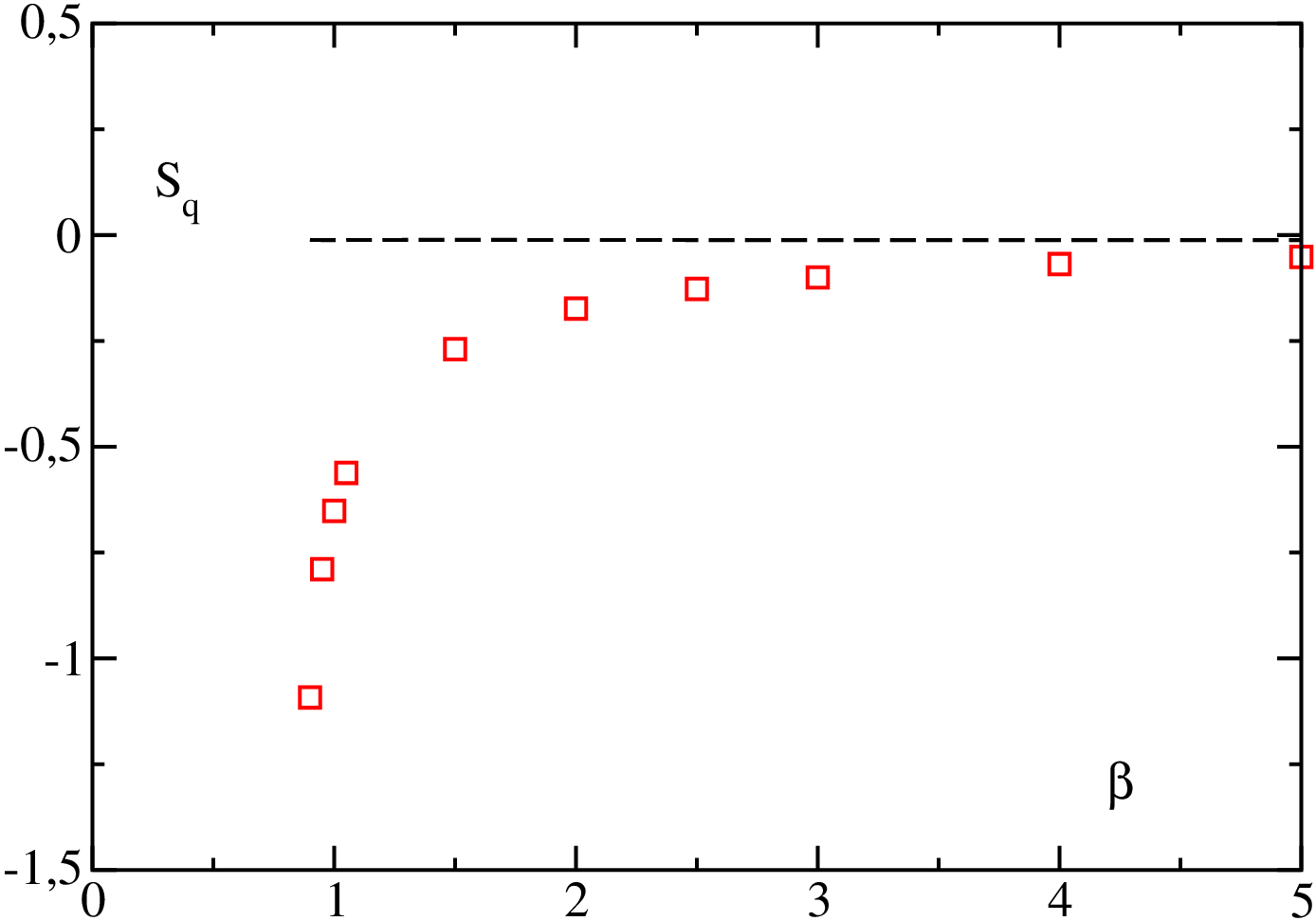}
\\\vspace{8mm}
(b)\hspace{10mm}\includegraphics[scale=0.28]{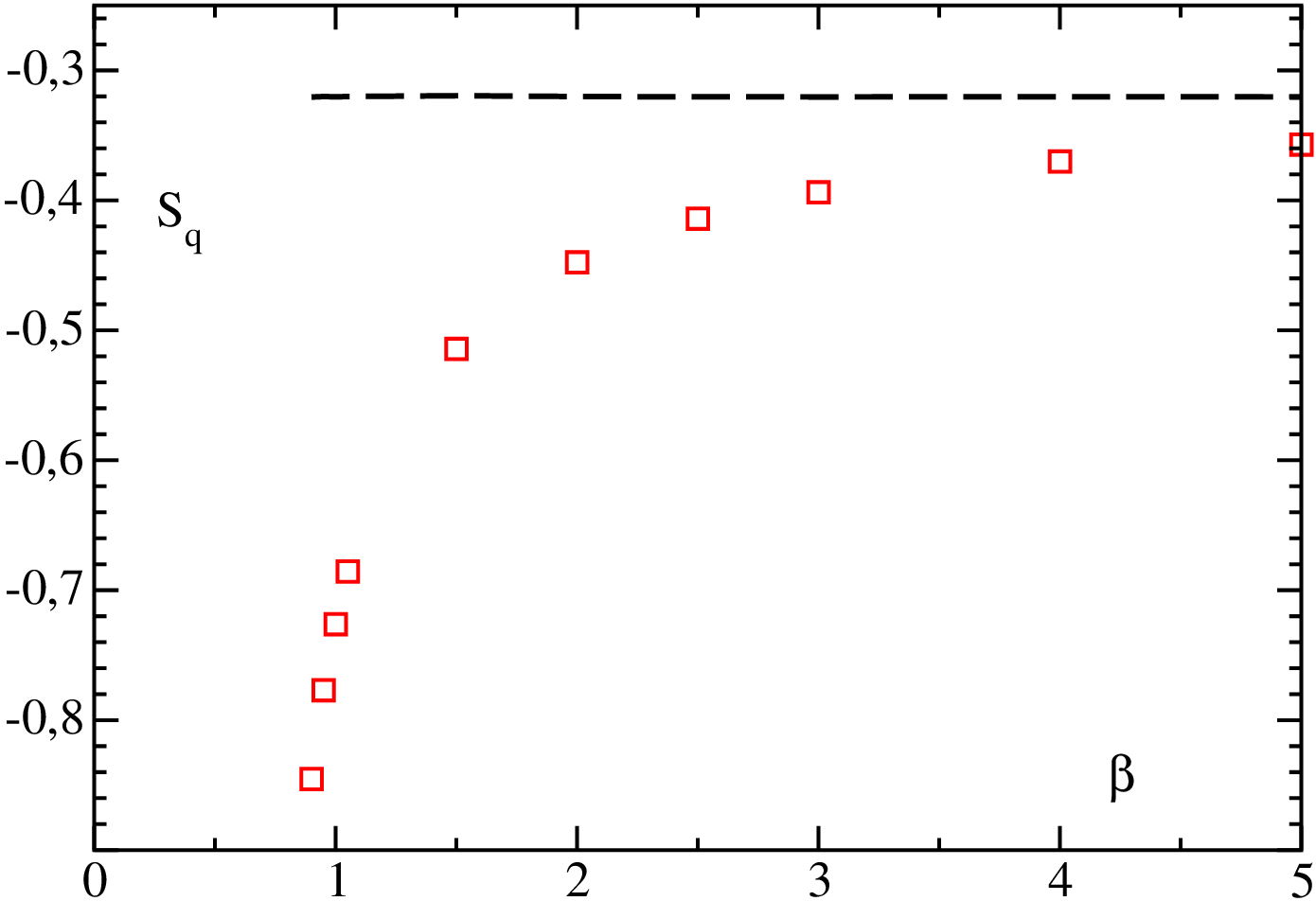}
\\\vspace{8mm}
(c)\hspace{10mm}\includegraphics[scale=0.28]{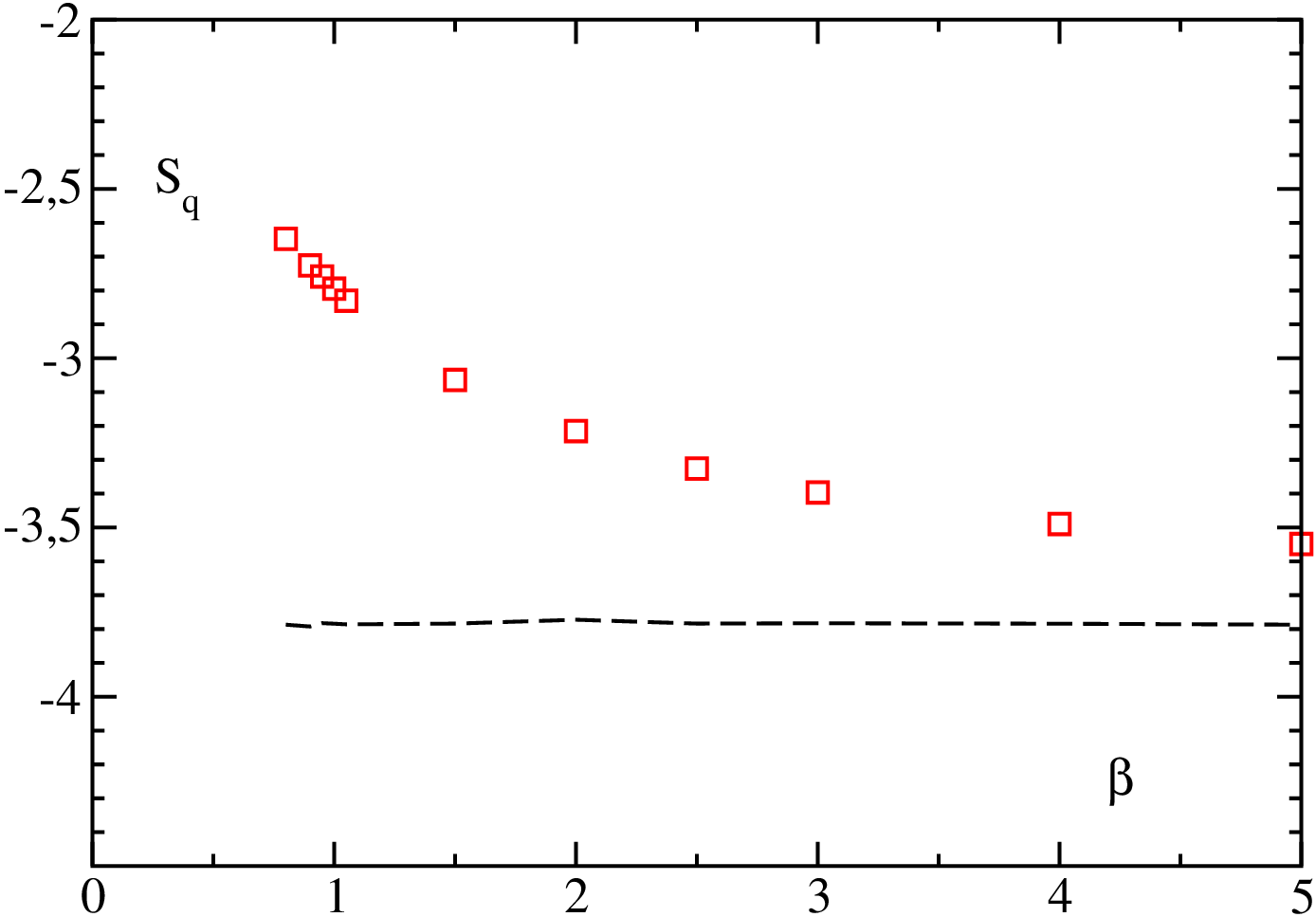}
\\\vspace{5mm}
  \caption{Quantum parts of the effective action as
 functions of $\beta$. Dashed
lines: one-loop approximation; squares: Hartree approximation; 
(a): $\alpha=0.2$; (b): $\alpha=0.5$; (c): $\alpha=0.9$.}
  \label{fig:betadependence}
\end{figure}

\begin{figure}[htbp]
\centering
(a)\hspace{10mm}\includegraphics[scale=0.32]{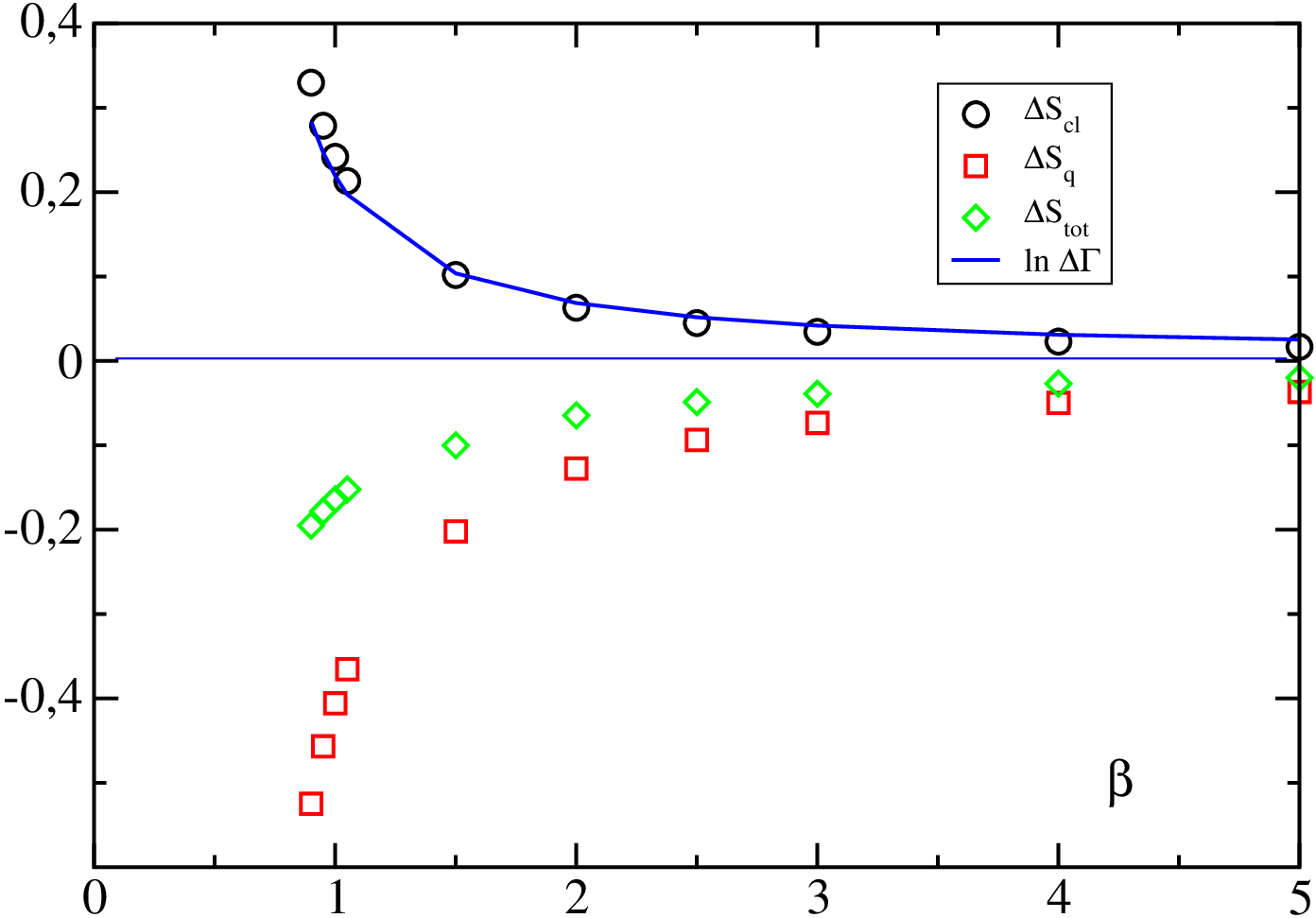}
\\\vspace{8mm}
(b)\hspace{10mm}\includegraphics[scale=0.32]{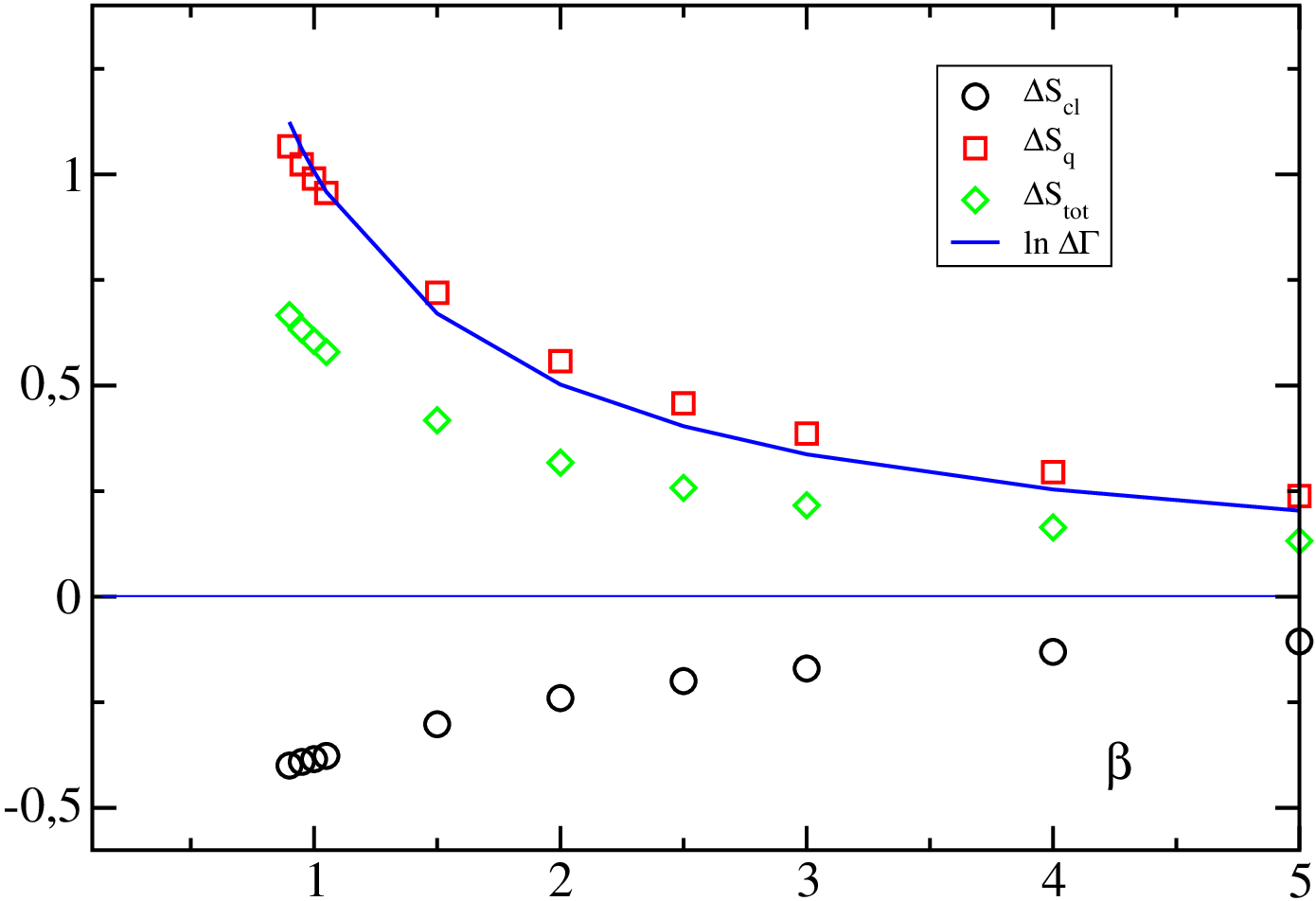}
\\\vspace{5mm}
  \caption{Differences between one-loop and Hartree approximation
of various parts of the effective action as functions of $\beta$.
Notations as specified in the text; 
(a): $\alpha=0.5$; (b): $\alpha=0.9$. }
  \label{fig:deltas}
\end{figure}

In Figs. \ref{fig:betadependence}(a-c) we plot the quantum
actions 
\be
S_{\rm q}^{1-\rm{loop}}
=\frac{1}{2}{\ln\det}'\frac{-\Delta_2+\calm^2}{-\Delta_2+m^2}\kma
\ee
as obtained in the one-loop approximation and
\be
S_{\rm q}^{\rm Hartree}=\frac{1}{2}{\ln\det}'\frac{-\Delta_2+\calm^2}
{-\Delta_2+m^2}-\frac{3\lambda}{8}\int d^2x \Delta^2(\vec x)\kma
\ee
obtained in the Hartree approximation as functions of $\beta$. 
It is found that the one-loop action,
which is independent of $\beta$, 
and $S_{\rm q}^{\rm Hartree}$ approach each other for large values of
$\beta$. When $\beta$ decreases 
below $\beta=1$ the difference increases rapidly, pointing,
for $\alpha=0.2$ and $0.5$,
to a kind of singularity at values of $\beta$ between
$0.8$ and $0.9$. Below this range the iteration ceases
to converge. During the iteration the value of $\phi(0)$
decreases, and the classical profile displays a minimum at some
finite value of $r$. 
The lack of convergence for $\beta \lesssim 0.8$ is not well-understood.
For some parameter sets, e.g., $\alpha=0.5$, Fig. 
\ref{fig:betadependence}(b), there seems to be a
kind of singularity of the quantum action, which, however, is
compensated by an opposite behavior in the classical action,
so that is not apparent in their sum, as can be seen in
Fig.\ref{fig:deltas}(a).
Some analytic idea, like reshuffling contributions between
the quantum and classical parts, could possibly cure the problem.
Indeed in a self-consistent scheme there  is no real
``classical'' part, as the profile depends on the 
quantum corrections.

\begin{figure}[ht]
\vspace{10mm}
  \centering
   \includegraphics[scale=0.35]{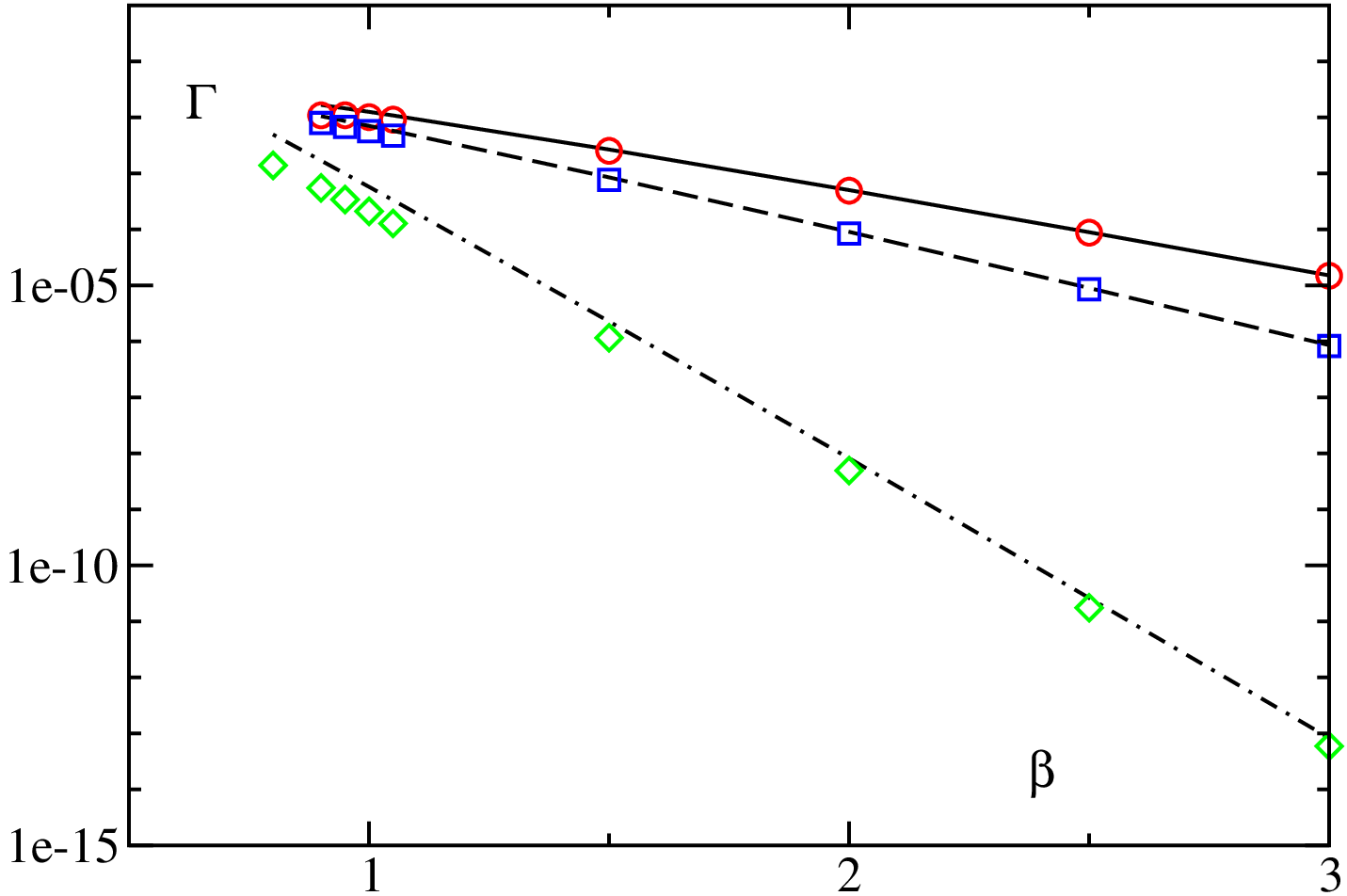}
\vspace{6mm}
  \caption{The transition rates in the one-loop and Hartree
approximations as functions of $\beta$.
One-loop approximation:
straight line: $\alpha=0.2$; dashed line: $\alpha=0.5$;
dash-dotted line:  $\alpha=0.9$; Hartree results:
circles: $\alpha=0.2$; squares: $\alpha=0.5$;
diamonds: $\alpha=0.9$.} 
 \label{fig:gammas}
\end{figure}

In Figs. \ref{fig:deltas}(a,b) we display the difference
between the one-loop and the Hartree approximations for various
quantities, for $\alpha=0.5$ and $\alpha=0.9$.
 $\Delta S_{\rm q}=S_{\rm q}^{\rm Hartree}-S_{\rm q}^{1-\rm{loop}}$, 
the difference of the
quantum actions, is found to be negative for $\alpha=0.5$
(and also for $\alpha=0.2$), but positive for $\alpha=0.9$. 
We also display the difference between the classical actions 
 $\Delta S_{\rm cl}=S_{\rm cl}^{\rm Hartree} -S_{\rm cl}^{1-{\rm loop}}$
and the total difference $\Delta S_{\rm tot}=\Delta S_{\rm cl}+
\Delta S_{\rm q}$. We see that even this
total difference has different signs for $\alpha=0.5$ and $\alpha=0.9$.
We should keep in mind again that the separation between classical
and quantum action is not unique and that the classical profile,
and therefore the prefactor, depends on the quantum fluctuations.
The total width, {\em which includes in the
prefactor the normalization of the zero mode} is for all $\alpha$ 
found to be  suppressed
in the Hartree approximation with respect to the one-loop
approximation. This is also displayed  in Figs. \ref{fig:deltas}(a) and 
(b),
where the solid line represents 
\be
\Delta \log \Gamma =-\log\frac{\Gamma^{\rm Hartree}}
{\Gamma^{1-{\rm loop}}}
\kma\ee
with the one-loop transition rate given by Eq. \eqn{Gammaoneloop},
and rate in the Hartree approximation given by Eq. \eqn{GammaHartree}.
This difference is positive for both $\alpha=0.5$ and $\alpha=0.9$.
These rates are displayed separately in Fig. \ref{fig:gammas}.

We have mentioned above that for the transition rates we use
the normalization of the translation mode as prefactor, as 
the virial theorem used in the one-loop approximation is no longer 
valid in the Hartree
approximation. In the one-loop approximation the normalization
of the zero mode and the classical action agree better than 
four significant digits, in the Hartree approximation 
the classical and total actions approach the normalization
of the zero mode for large $\beta$, but $\beta\simeq 1$ both of them
differ from it by factors up to $3$. 

\begin{figure}[ht]
\vspace{10mm}
  \centering
   \includegraphics[scale=0.35]{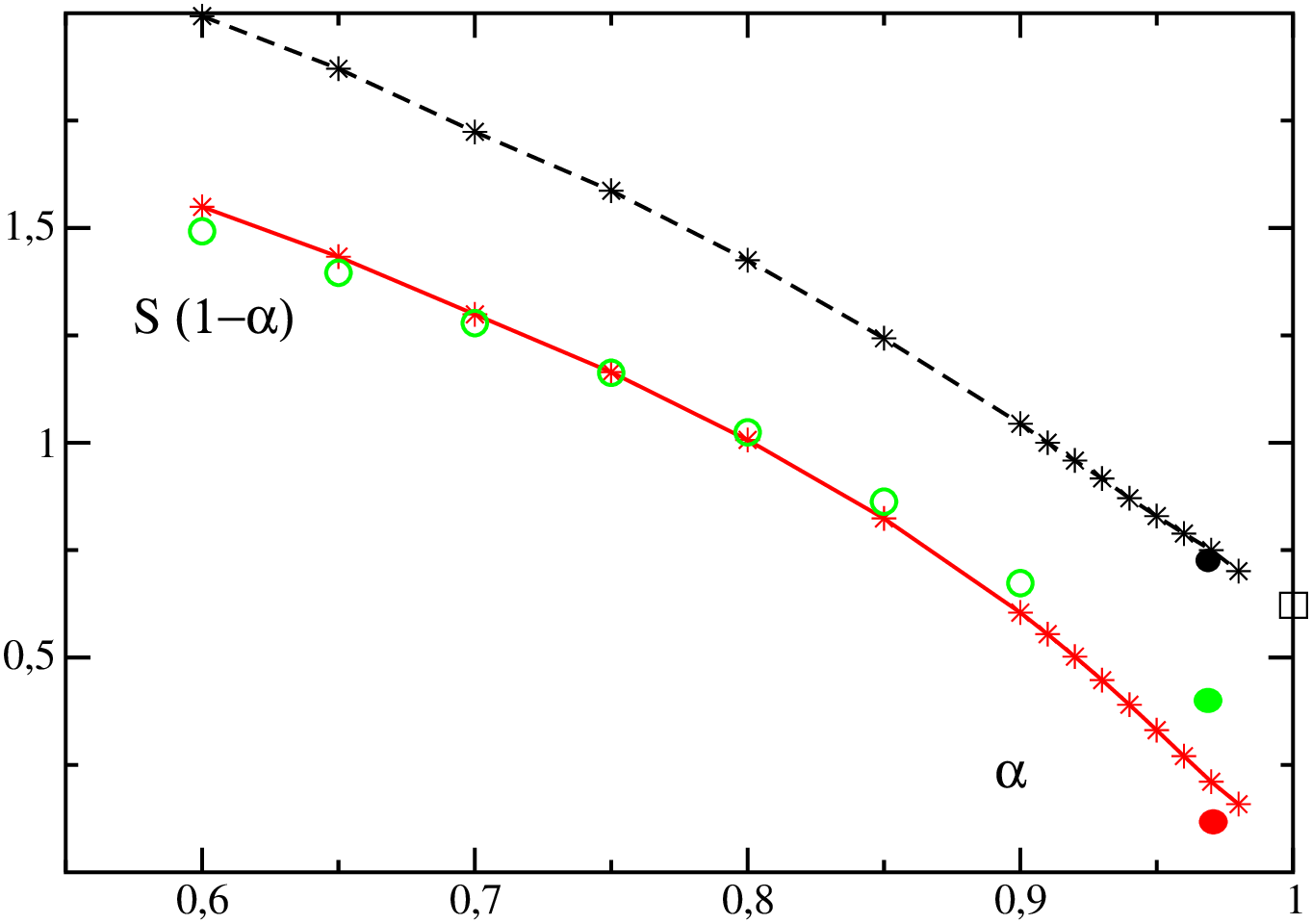}
\vspace{6mm}
  \caption{Comparison with Ref. \cite{Bergner:2003id}:
Dashed line with stars: our results for $S_{\rm cl}^{1-{\rm loop}}$;
full line with stars: our results for $S_{\rm tot}^{1-{\rm loop}}$;
empty circles: our results for $S^{\rm H}_{\rm tot}$; full
circles: the corresponding results of Ref. \cite{Bergner:2003id};
empty square: thin wall limit of the classical action; all
actions are multiplied with $(1-\alpha)$.
 }
\label{fig:comparison}
\end{figure}

As we have mentioned above the procedure becomes unstable for
 $\beta\lesssim 1$; the iteration ceases to converge. 
Bergner and Bettencourt \cite{Bergner:2003id}, on the other hand
present results for $m=1.3469$, and, in our conventions
\footnote{The potential of \cite{Bergner:2003id}
is given by $V(\phi)=\lambda(\phi^2-v^2)^2/4-\epsilon(\phi+v)/(2v)$.
In order to compare it with ours it has to be shifted 
in $\phi$ such that the left minimum
is at $\phi=0$.} $\lambda=2$
and $\eta=0.9685$. They use the same renormalization convention
as we do. Their parameters  correspond to $\alpha=0.967$ and $\beta=0.877$.
This is close to the thin wall limit,
and at a value of $\beta$ for which our procedure
to becomes unreliable. The features of the solution
found in Ref. \cite{Bergner:2003id} are quite different from
those found here: there the eigenvalue spectrum
displays neither a zero nor an unstable mode.
It may be that the tunnelling in this region,
$\beta \lesssim 0.9,\alpha > 0.9$
takes place in a qualitatively different way. If this is
so then our procedure certainly breaks down on more than
numerical reasons as it is based on an conventional treatment
of zero and unstable modes. In order to obtain a comparison 
of the result of Ref. \cite{Bergner:2003id} with ours, we have
determined the various actions for $\beta=0.877$ and 
$m=1.3469$ fixed, as a function of $\alpha$. These results
are plotted in Fig. \ref{fig:comparison}. 
The classical action of Ref.. \cite{Bergner:2003id}
agrees reasonably well with our results.
We also show the thin wall limit, Eq. \eqn{thinwall}. 
The total one-loop action
can be computed up to $\alpha=0.98$. The total one-loop
action obtained in Ref. \cite{Bergner:2003id} is somewhat
lower than our values. Surprisingly our
results for the Hartree approximation seem to extrapolate quite 
naturally towards the result of Ref.\cite{Bergner:2003id}.

\section{Conclusions and outlook}
\label{conclusion}
\setcounter{equation}{0}
We have presented here a calculation of the transition rate
for the false vacuum decay in $(1+1)$ dimensions in the Hartree
approximation. We have used analytic and numerical methods that 
have been applied  previously to other systems, for the
computations of functional determinants 
\cite{Baacke:1993ne,Baacke:1991nh,Baacke:2003uw} 
and zero point energies \cite{Baacke:1991nh,Baacke:1998nm}. 
Both techniques are based on
 mode functions of the fluctuation operator at $\nu^2=0$,
 as they appear in the
standard representation of Green's functions. Here these
techniques are used for the first time in conjunction; as one sees
this leads to a rather simple and effective computation scheme
with various cross checks. We should like to add, that these
techniques are well-suited for coupled-channel systems as well
\cite{Baacke:1995bw}. Renormalization can be dealt with in exact
correspondence to standard perturbation theory; 
this feature extends to applications in
higher space-time dimensions and, allowing for
dimensional regularization, to computations  in gauge theories. 

We have presented results for a substantial part of parameter space,
using the parameters $\beta =m^4/4\eta^2$  and $\alpha=\lambda\beta/m^2$.
Bounces exist only for $\alpha<1$, $\alpha=1$ being the
thin wall limit. The second parameter $\beta$ weighs the 
ratio between classical and quantum contributions to the effective action.
We find that for $\beta \gtrsim 0.8$, and for $\alpha \lesssim 0.9$
our iterative determination of the field configuration in the
Hartree approximation converges. The unstable mode and
an almost-zero mode in the $p$-wave, identified as the translation mode,
persist during the iteration and are handled as in the
one-loop approximation.
We find that the transition rates are generally suppressed in the
Hartree approximation, as compared to the one-loop rates. However
the corrections are relatively small, less than an order of
magnitude. This has been found similarly for the one-loop
quantum corrections for instanton transitions in
the two-dimensional Abelian Higgs model \cite{Baacke:1994bk} and may
be an general feature of two-dimensional models.

The only similar computation which is available
at present is the one of Ref. \cite{Bergner:2003id}.
Unfortunately our procedure does not converge for the
parameters chosen by these authors, which in our convention
are $\alpha=0.967$ and $\beta=0.877$. Their self-consistent
configuration is qualitatively different from ours in that the fluctuation
operator has no unstable and translation modes. If this
is the case in this part of the parameter space then
it means that the tunneling proceeds in a qualitatively different
way, there. Nevertheless we have found that our Hartree results seem
to extrapolate naturally towards the ones of Ref. \cite{Bergner:2003id}. 

The methods used here for the computation of self-consistent
solutions for bounces naturally extends to bounces in 
higher space dimensions and to other classical solutions
in quantum field theory. In particular, the techniques for
covariant regularization and renormalization are available
\cite{Baacke:1991nh,Baacke:1993ne} in the framework we have established here.

\section{Acknowledgments}
N. K.  was supported by the \emph{Deutsche Forschungsgemeinschaft} 
as a member of \emph{Gra\-du\-ier\-ten\-kol\-leg 841}.

\bibliography{bounce}
\bibliographystyle{h-physrev4}
\end{document}